\pdfoutput=1

\documentclass[final,1p,authoryear]{elsarticle}

\usepackage{amssymb}

\usepackage{amsmath,amsfonts}
\usepackage{todonotes}
\usepackage{booktabs}
\usepackage{siunitx}
\usepackage{subfig}
\usepackage{hyphenat}
\usepackage{xurl}
\usepackage[breaklinks=true]{hyperref}
\usepackage{textpos}

\usepackage{todonotes}

\makeatletter
\def\ps@pprintTitle{   \let\@oddhead\@empty
   \let\@evenhead\@empty
   \def\@oddfoot{\reset@font\hfil\thepage\hfil}
   \let\@evenfoot\@oddfoot
}
\makeatother

\graphicspath{{./img/}}

\begin{document}

\begin{textblock*}{190mm}(-3cm,-5cm)
\noindent \footnotesize The peer-reviewed version of this paper is
published in Computers and Education: Artificial Intelligence at
\url{https://doi.org/10.1016/j.caeai.2021.100014}.
This version is typeset by the author and differs only in pagination and typographical detail.
\end{textblock*}

\begin{frontmatter}

\title{ColorShapeLinks: A board game AI competition for educators and students}

\author{Nuno Fachada}

\address{Lusófona University, COPELABS / HEI-Lab\\
Campo Grande, 376, Lisbon, Portugal\\
\ead{nuno.fachada@ulusofona.pt}}

\begin{abstract}
ColorShapeLinks is an AI board game competition framework specially designed for
students and educators in videogame development, with openness and accessibility
in mind. The competition is based on an arbitrarily-sized version of the
Simplexity board game, the motto of which, ``simple to learn, complex to
master'', is curiously also applicable to AI agents. ColorShapeLinks offers
graphical and text-based frontends and a completely open and documented
development framework built using industry standard tools and following
software engineering best practices. ColorShapeLinks is not only a competition,
but both a game and a framework which educators and students can extend and use
to host their own competitions. It has been successfully used for running
internal competitions in AI classes, as well as for hosting an international
AI competition at the IEEE Conference on Games.
\end{abstract}

\begin{keyword}
Board games \sep Videogame development \sep Computer games \sep Simplexity
\sep AI competition \sep AI agent \sep AI curriculum \sep DotNet

\end{keyword}

\end{frontmatter}

\section{Introduction}
\label{sec:intro}

ColorShapeLinks is an artificial intelligence (AI) competition framework for
the Simplexity board game \citep{simplexitybgg} with arbitrary dimensions. It is
a similar game to Connect-4, with pieces defined not only by color, but also
by shape.

The ColorShapeLinks development framework offers Unity \citep{unity3d} and .NET
console frontends. Agents are implemented in C\# and run unmodified in either
frontend. Unity and C\# are widely used in the games industry
\citep{toftedahl2019taxonomy} and for game development education
\citep{dickson2015using,dickson2017an,comber2019engaging,fachada2020topdown,hmeljak2020developing},
making the competition especially accessible to this audience. The framework is
open source, fully documented and developed following best practices in software
engineering, allowing it to be studied and extended by educators, researchers
and students alike \citep{lakhan2008open,jimenez2017four}. Furthermore, it
contains all the tooling for setting up competitions. It can be used for
internal competitions in AI courses, for example, or for running international
AI competitions, as demonstrated in the 2020 edition of the IEEE Conference on
Games (IEEE CoG). In this regard, ColorShapeLinks is not simply a software tool
that uses AI technologies for educational purposes \citep{chen2020multi}, but a
software toolkit to both teach and learn board game AI.

This paper is organized as follows. In Section~\ref{sec:background}, the state
of the art in board game AI---and how ColorShapeLinks fits in---is briefly
discussed. The ColorShapeLinks board game and its original version, Simplexity,
are characterized in Section~\ref{sec:csl}. In Section~\ref{sec:motivation}, the
motivation for developing ColorShapeLinks, as well as the educational context in
which the framework took form, are examined. The development framework, namely
its architecture, existing frontends, agent implementation, included agents, and
availability, are reviewed in Section~\ref{sec:devframework}.
In Section~\ref{sec:deployments}, we describe how ColorShapeLinks was used to
host two internal competitions in an AI course unit, as well as a fully-fledged
international AI competition in the IEEE CoG 2020 conference. The implications
and limitations of the framework and of the reported outcomes are discussed in
Section~\ref{sec:limitations}. The paper closes with
Section~\ref{sec:conclusions}, in which we present some conclusions.

\section{Background}
\label{sec:background}

Computer programs for playing classical board games such as Chess, Draughts or
Go, were the first known application of AI for games \citep{millington2019ai}.
AI research in games has subsequently grown far beyond the domain of board
games \citep{yannakakis2018artificial}. Nonetheless, these continue to be a
focus of active research, not only in AI techniques for playing particular
games---e.g., Go \citep{silver2016mastering}---but also for playing board games
in general \citep{konen2019general,kowalski2019regular,piette2019ludii}, or even
creating new ones \citep{stephenson2019ludii,kowalski2016evolving}. Furthermore,
new board game AI challenges and competitions continue to be proposed
\citep{justesen2019blood}.

While board games are probably one of the easiest ways to introduce AI for games
to students \citep{drake2011teaching,chesani2017game}, state-of-the-art board
game AI research is gaining some distance from both industry and education in
videogame development. Requirements such as general game playing capabilities
for the AI or knowledge of general game specification languages
\citep{stephenson2019ludii, kowalski2019regular} can potentially raise the entry level for newcomers and/or discourage
prospective participants which could otherwise bring new ideas to academia---or
at least get involved in its processes.

ColorShapeLinks aims to reduce this gap by offering an approachable, open and
flexible AI competition for the videogame development education audience. This
is accomplished by making the development of an AI agent very simple (e.g.,
write a single method and test it in Unity), while allowing for considerable
customization via a modular framework architecture. This same modularity enables
games and full tournaments to run within and without Unity. While accessible for
undergraduates and non-specialist game developers, ColorShapeLinks provides a
challenging and intricate board game competition, addressable with vastly
different techniques, from knowledge-based methods \citep{allis1988knowledge} to
machine learning techniques \citep{silver2016mastering}, or anything in between.

It should be noted that ColorShapeLinks could be implemented under a general
game playing system such as RBG \citep{kowalski2019regular} or Ludii
\citep{piette2019ludii}, and this would probably be an excellent exercise.
However, using these frameworks would mean losing some of the educational
advantages offered by ColorShapeLinks, namely accessibility, openness (e.g.,
Ludii was not open source at the time of writing) and keeping the focus on
general game design and development education while providing students with a
broad AI for games background.

\section{ColorShapeLinks}
\label{sec:csl}

ColorShapeLinks is a version of the Simplexity board game with arbitrary
and parameterizable dimensions. We describe Simplexity and
ColorShapeLinks in Subsections~\ref{sec:csl:simplexity} and
\ref{sec:csl:unbounded}, respectively, highlighting their computational
characteristics in Subsection~\ref{sec:csl:characteristics}.

\subsection{The Simplexity Board Game}
\label{sec:csl:simplexity}

\begin{figure}[tb]
    \centering
    \subfloat[White wins with white pieces.\label{fig:victory:whitewhite}]{
        \includegraphics[width=0.5\textwidth]{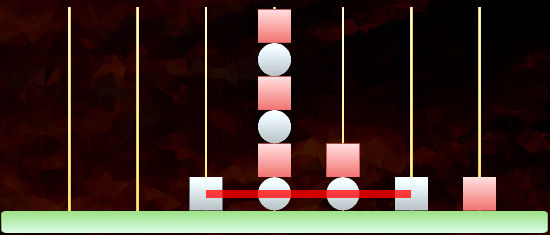}}
    \subfloat[White wins with round pieces.\label{fig:victory:whiteround}]{
        \includegraphics[width=0.5\textwidth]{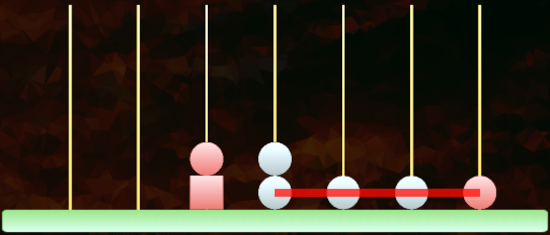}}\\

    \subfloat[Red wins with red pieces.\label{fig:victory:redred}]{
        \includegraphics[width=0.5\textwidth]{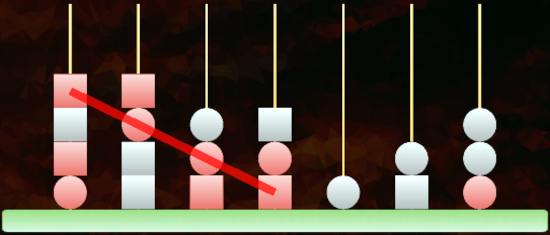}}
    \subfloat[Red wins with square pieces.\label{fig:victory:redsquare}]{
        \includegraphics[width=0.5\textwidth]{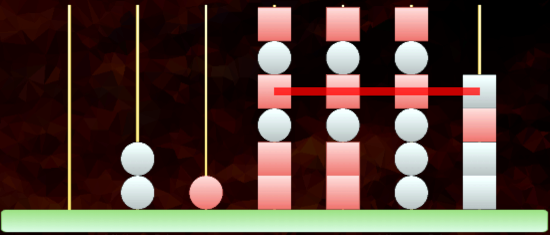}}

    \caption{Possible victory conditions in ColorShapeLinks using standard
        Simplexity rules.}
    \label{fig:victory}
\end{figure}

The Simplexity board game is similar to Connect-4 in two regards:
1) the first player to connect four pieces of a given type wins; and,
2) the base board size is $6 \times 7$. The crucial difference is that, in
Simplexity, pieces are defined not only by color, but also by shape---round or
square. Player 1, \textit{white}, wins if it can connect either four
round pieces or four white pieces. Likewise for player 2, \textit{red}, but
with square or red pieces. Players begin with 21 pieces of their color, 11
of which square, and the remaining 10 round. The catch here is that shape has
priority over color as a winning condition. Therefore, a player can lose in its
turn when placing a piece of the opponent's winning shape.
Table~\ref{tab:simplexity} summarizes these rules and
\figurename~\ref{fig:victory} shows the possible winning conditions.

\begin{table}[tb]        \centering
        \caption{Simplexity rules.}
    \label{tab:simplexity}
    \begin{tabular}{lll}
        \toprule
    & \bfseries Player 1 & \bfseries Player 2\\
        \midrule
    Plays first?         & Yes & No\\
    Plays with           & White pieces & Red pieces\\
    Begins with          & $11\times$ white square pieces & $11\times$ red square pieces\\
                         & $10\times$ white round pieces & $10\times$ red round pieces\\
    Wins with line of    & $4\times$ round pieces & $4\times$ square pieces\\
    (shape has priority) & $4\times$ white pieces & $4\times$ red pieces\\
        \bottomrule
    \end{tabular}
\end{table}

\subsection{The ColorShapeLinks Board Game}
\label{sec:csl:unbounded}

ColorShapeLinks is a Simplexity game parameterizable with respect to board
dimensions, number of pieces required for a winning sequence, and initial
number of round and square pieces. Table~\ref{tab:cslparams} shows the
available parameters as well as the symbols used for them in the remainder
of this paper.

\begin{table}[tb]
            \centering
    \caption{ColorShapeLinks parameters.}
    \label{tab:cslparams}
    \begin{tabular}{clr}
        \toprule
    \bfseries Symbol & \bfseries Name & \bfseries Default\textsuperscript{a}\\
        \midrule
    $r$ & Rows                     & 6\\
    $c$ & Columns                  & 7\\
    $w$ & Win sequence             & 4\\
    $s$ & Square pieces per player & 11\\
    $o$ & Round pieces per player  & 10\\
        \bottomrule
    \multicolumn{3}{l}{\mbox{ }\textbf{\textsuperscript{a}} i.e.,
        a regular game of Simplexity.}
    \end{tabular}
\end{table}

\subsection{Characteristics}
\label{sec:csl:characteristics}

ColorShapeLinks can be characterized as follows
\citep{yannakakis2018artificial,millington2019ai}:

\begin{itemize}
    \item It is a \textbf{turn-based} game.
    \item It is a \textbf{two-player}, \textbf{zero-sum adversarial} game.
    \item It is \textbf{deterministic}, i.e., there is no random element
        influencing the game state.
    \item It has \textbf{perfect information}, i.e., the board state is fully
        observable.
    \item The \textbf{maximum number of turns} is given by
        \begin{equation*}
            t_\text{max} = \min{\{r \cdot c, 2(s + o)\}}
        \end{equation*}
        i.e., it is equal to
        the minimum between the number of board positions, $r \cdot c$, and
        the total pieces available to be played, $2(s + o)$.
    \item Since each board position can be in one of five states (empty, white
        circle, white square, red circle or red square), an upper bound for
        the \textbf{state space} is given by $5^{t_\text{max}}$. In practice the
        state space will be considerably smaller, since this value includes
        invalid states, for example when pieces are on top of empty cells.
    \item The initial \textbf{branching factor} is $2 \times c$ (2 shapes, $c$
        columns), although it may decrease during the game as columns are
        filled and pieces of a certain shape are played-out. Consequently, an
        upper bound for the game tree size is given by
        $(2 \cdot c)^{t_\text{max}}$.
\end{itemize}

These characteristics place ColorShapeLinks in an interesting position for
AI research. The game, even with standard Simplexity parameters, cannot
currently be solved using a brute force approach in a short amount of time. Like
most games nowadays, machine learning techniques (e.g., deep learning) together
with a tree search approach such as Monte Carlo Tree Search (MCTS)
\citep{coulom2006efficient}, for example, will certainly be able to produce
hard-to-beat agents. However, the limited and well-defined ruleset leaves the
door open for knowledge-based or even analytical solutions.

\section{Educational Context and Motivation}
\label{sec:motivation}

ColorShapeLinks was originally developed as a Unity-only assignment for an AI
for Games course
unit\footnote{\url{https://github.com/VideojogosLusofona/ia_2019_board_game_ai}}
at Lusófona University's Videogames BA. This is an evenly interdisciplinary
degree \citep{mateas2007design}, meaning that while it possesses solid Computer
Science (CS) fundamentals, it is more limited in this regard than
technology-focused curriculums, giving equal ground to Game Design
and Art courses. The Unity game engine is used in most of the course units,
since it is easy to learn, free, cross-platform, and widely used in education and
actual game development
\citep{dickson2015using,dickson2017an,comber2019engaging,toftedahl2019taxonomy,fachada2020topdown,hmeljak2020developing}.
Consequently, and to increase student's proficiency with Unity, the engine's
scripting language, C\#, is lectured independently in two programming course
units \citep{fachada2020topdown}. The AI for Games course unit threads this fine
line by implementing a hands-on, Unity and industry-oriented program based on
selected parts of Millington's \textit{AI for Games} textbook
\citep{millington2019ai}. The course unit addresses topics such as heuristics,
board games, movement, pathfinding, decision making, learning and procedural
content generation. However, more complex and/or academic materials are avoided.
With respect to the topic of board games, the course focuses on the Minimax
algorithm, as well as several variations and optimizations, such as alpha-beta
pruning, move ordering and iterative deepening. Techniques such as
transposition tables and MCTS are not currently lectured, although in the
specific case of MCTS this might change given the proven usefulness of the
method in a variety of different games and scenarios
\citep{yannakakis2018artificial}.

Due to its wide-ranging design, Lusófona's Videogames degree attracts students
from various areas of study and with substantially different interests. An issue
with this type of curriculum design is that it can be difficult to motivate
students enrolled in courses outside their main preferences. This is often the
case of art and design-inclined students in CS courses in general
\citep{fachada2018db,fachada2019desafios,deandrade2020fun}, and AI in
particular \citep{lin2021modeling}. Thus, student motivation strategies become
particularly relevant in this context \citep{pintrich2003motivational}. Board
game competitions in AI courses have been shown to motivate students in
studying course materials and in autonomously search for solutions beyond the
scope of the course's program and lectures \citep{chesani2017game}. Hence, and
with the goal of increasing motivation among students---particularly those whose
primary interests do not lie in CS---a board game AI assignment featuring an
internal classroom competition was devised.
Given the variety of student backgrounds, the project had to be made accessible
for everyone, while challenging for more advanced and/or CS-oriented students.
A deterministic, fully observable two-player, non-trivial board game was an
obvious choice. Another requirement was that it was not a very well-known game,
so that not much AI code was available online, forcing students to be original.
Simplexity \citep{simplexitybgg} ticks those boxes, as it was only used once in
an AI course \citep{wilkins2012simplexity}, to our knowledge. Furthermore,
Simplexity was implemented as a console C\# project (no AI) two years
prior\footnote{\url{https://github.com/VideojogosLusofona/lp1_2017_p1}},
thus being a perfect choice, since students already knew the base game. This led
to the development of ColorShapeLinks, an AI assignment and competition, first
introduced in the first semester of the 2019/20 academic year.

Moreover, some authors have reported positive educational outcomes when creating
assignments based on international AI competitions
\citep{kim2013game,yoon2015challenges}. They argue that, by offering students
the possibility of submitting their solutions to international events, and of
comparing their solutions with the state of the art, their engagement and
motivation are improved. Following this line of reasoning,
ColorShapeLinks was proposed as a competition for the IEEE CoG 2020 conference.
Since the first version of the framework was only suitable for basic classroom
competitions, it was extended to support more advanced use cases. Additions
included a console mode, for advanced debugging and analysis of agents, and a
set of scripts for setting up automatically running tournaments. Thus,
ColorShapeLinks recast its role as an internal competition during the second
semester of the 2019/20 academic year, while at the same time hosting the
eponymous competition at the IEEE CoG 2020 conference. These experiences are
described in Section~\ref{sec:deployments}.

Beyond the educational benefits of ColorShapeLinks discussed thus far, and the
fact that Unity is widely used in education and industry, it should also be
noted that: (1) C\# is used as a scripting language in several major game engines
\citep{unity3d,xenko,cryengine,godotengine,monogame,flaxengine,unigine,waveengine};
and, (2) neither Unity nor C\# are very common in game AI competitions. Thus,
ColorShapeLinks has the potential to more generally reduce the gap
between academic AI and game development industry/education. This is especially
relevant for interdisciplinary and/or industry-oriented programs such as
Lusófona's Videogames BA.

\section{Development Framework}
\label{sec:devframework}

The ColorShapeLinks development framework offers two application frontends,
one for the Unity game engine, the other a text-based affair aimed at
terminal use. This allows students familiar with Unity to start implementing
an agent from their comfort zone, advancing to the text-based console frontend
if Unity development limitations begin to show. Advanced students or researchers
can skip the Unity frontend altogether, and start with the console frontend from
the offset.
The framework architecture supporting this flexibility is described in
Subsection~\ref{sec:devframework:arch}.
Subsection~\ref{sec:devframework:frontends} details the two available frontends,
highlighting their advantages and disadvantages.
The basics of implementing an agent are discussed in
Subsection~\ref{sec:devframework:implementing}, while the agents included with
the framework are presented in Subsection~\ref{sec:devframework:agents}.
Finally, the framework's availability, documentation and licensing are outlined
in Subsection~\ref{sec:devframework:availability}.

\subsection{Architecture}
\label{sec:devframework:arch}

\begin{figure}[tb]    \centering
    \includegraphics[width=1\textwidth]{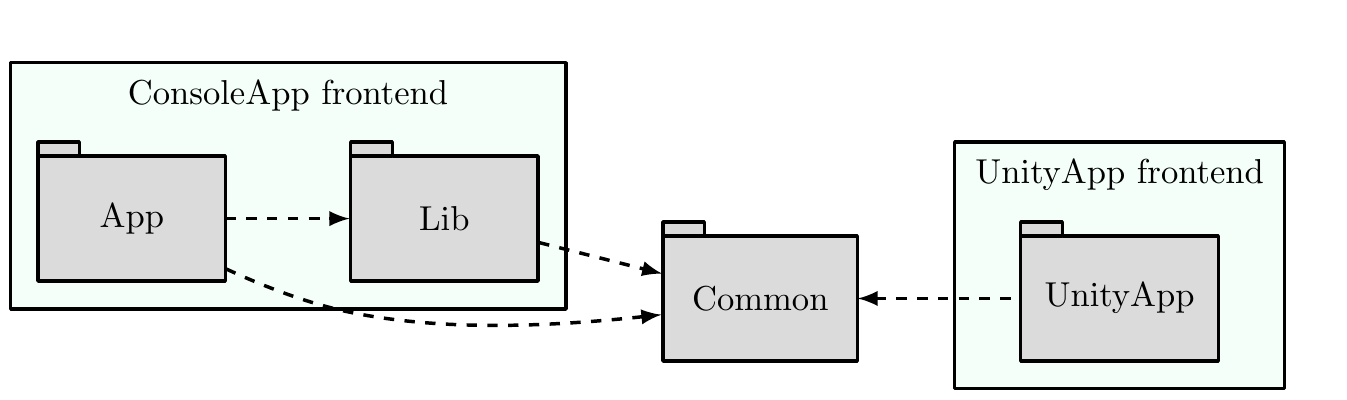}
    \caption{Internal organization of the ColorShapeLinks development
        framework. Arrows represent dependencies between separate .NET
        projects.}
    \label{fig:arch}
\end{figure}

The development framework is organized around three main components, namely
the Unity frontend (UnityApp, a single .NET/Mono project), the console
frontend (ConsoleApp, composed of two .NET projects), and the Common .NET
project. This organization is shown in \figurename~\ref{fig:arch}, and
discussed with additional detail in the following paragraphs.

\subsubsection{The Common Project}
\label{sec:devframework:arch:common}

The Common project is a .NET class library which constitutes the
core of the framework. It defines the fundamental models---from a
Model-View-Controller (MVC) perspective \citep{sarcar2020design}---of
the ColorShapeLinks game, such as the board, its pieces or performed moves,
and is a dependency of the remaining projects. It is further subdivided in the
\texttt{AI} and \texttt{Session} namespaces.
The former defines AI-related abstractions, such
as the \texttt{AbstractThinker} class, which AI agents must extend,
as well as a manager for finding and instantiating concrete AI agents. The
latter specifies a number of match and session-related interfaces, as well as
concrete match and session (i.e., tournament) models.

\subsubsection{The ConsoleApp Frontend Projects}
\label{sec:devframework:arch:consoleapp}

The ConsoleApp is composed of two .NET projects, App and Lib, both of which
depend on the Common class library, as shown in \figurename~\ref{fig:arch}.
The App project is a .NET console application with an internal dependency
on the Lib project, itself a .NET class library. The App
project provides the actual console frontend, namely the text user interface
(TUI) with which the user interacts in order to run ColorShapeLinks matches
and sessions.

The Lib class library acts as an UI-independent ``game engine'', offering
match and session controllers, as well as interfaces for the associated event
system, allowing to plug in renderers (views, in MVC parlance) or other
event handling code at runtime. It serves as a middleware between the Common
library and frontend applications, such as the one implemented in the App
project. It is not used by the Unity implementation, since Unity already
provides its own game engine logic, forcing match and session controllers to
be tightly integrated with its frame update cycle. Nonetheless, the Lib
class library makes the creation of new ColorShapeLinks TUIs or GUIs very
simple, as long as they are not based on highly prescriptive frameworks
such as Unity.

\subsubsection{The UnityApp Frontend Project}
\label{sec:devframework:arch:unityapp}

The UnityApp is a ColorShapeLinks frontend implemented in the Unity game
engine. Like the ConsoleApp, it is designed around the MVC design pattern,
making use of the models provided by the Common library. In this case, however,
the views and controllers are tightly integrated with the Unity engine.

\subsection{Frontends}
\label{sec:devframework:frontends}

The two available frontends, for Unity and console, have similar capabilities.
Both are capable of performing matches and tournaments involving multiple AI
agents and human players. An agent can be used without modification
when moving from one frontend to the other.
There are, however, advantages in using the console frontend. The following
paragraphs offer additional detail on each frontend.

\subsubsection{The Unity Frontend}
\label{sec:devframework:frontends:unity}

\begin{figure}[tb]
    \centering
    \includegraphics[width=1\textwidth]{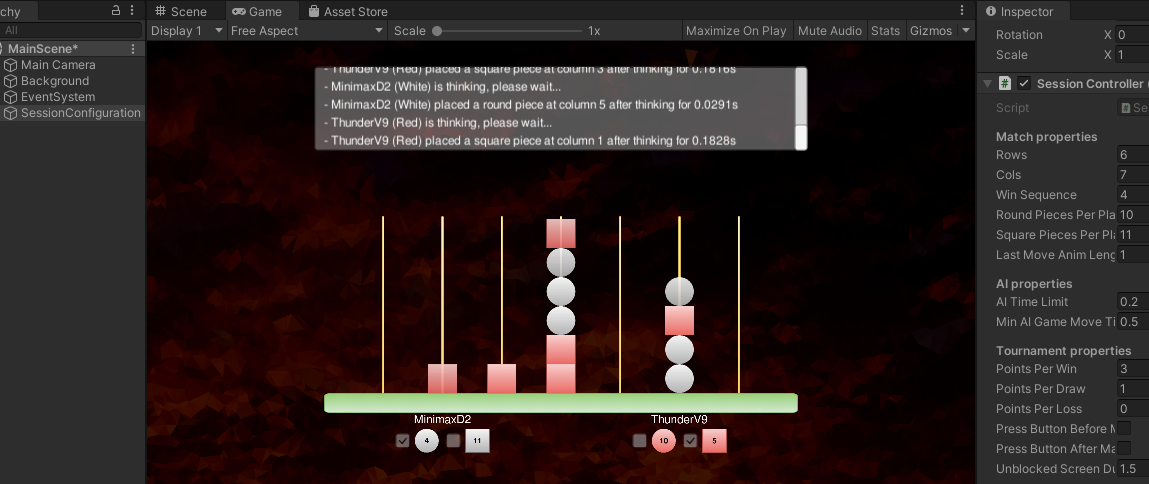}
    \caption{Running ColorShapeLinks using the Unity Editor UI.}
    \label{fig:unity}
\end{figure}

In the Unity frontend, matches and tournaments are played within the Unity
Editor, though it is possible to create a standalone build with prefixed
match or tournament configurations. The rationale behind this choice is that
ColorShapeLinks is at its core a development framework. Therefore, it makes
sense this is done within the Unity Editor, which is a development environment.
A game of ColorShapeLinks running within the Unity Editor is shown in
\figurename~\ref{fig:unity}.

Developing an agent within the Unity editor has the disadvantage that games run
considerably slower than in the console. Furthermore, constantly creating
standalone builds with prefixed configurations is not practical. Therefore,
while this frontend makes ColorShapeLinks approachable, it should be considered
more of an introduction to the competition than a definitive way of implementing
advanced state-of-the-art ColorShapeLinks agents.

\subsubsection{The Console Frontend}
\label{sec:devframework:frontends:console}

\begin{figure}[tb]    \includegraphics[width=1\textwidth]{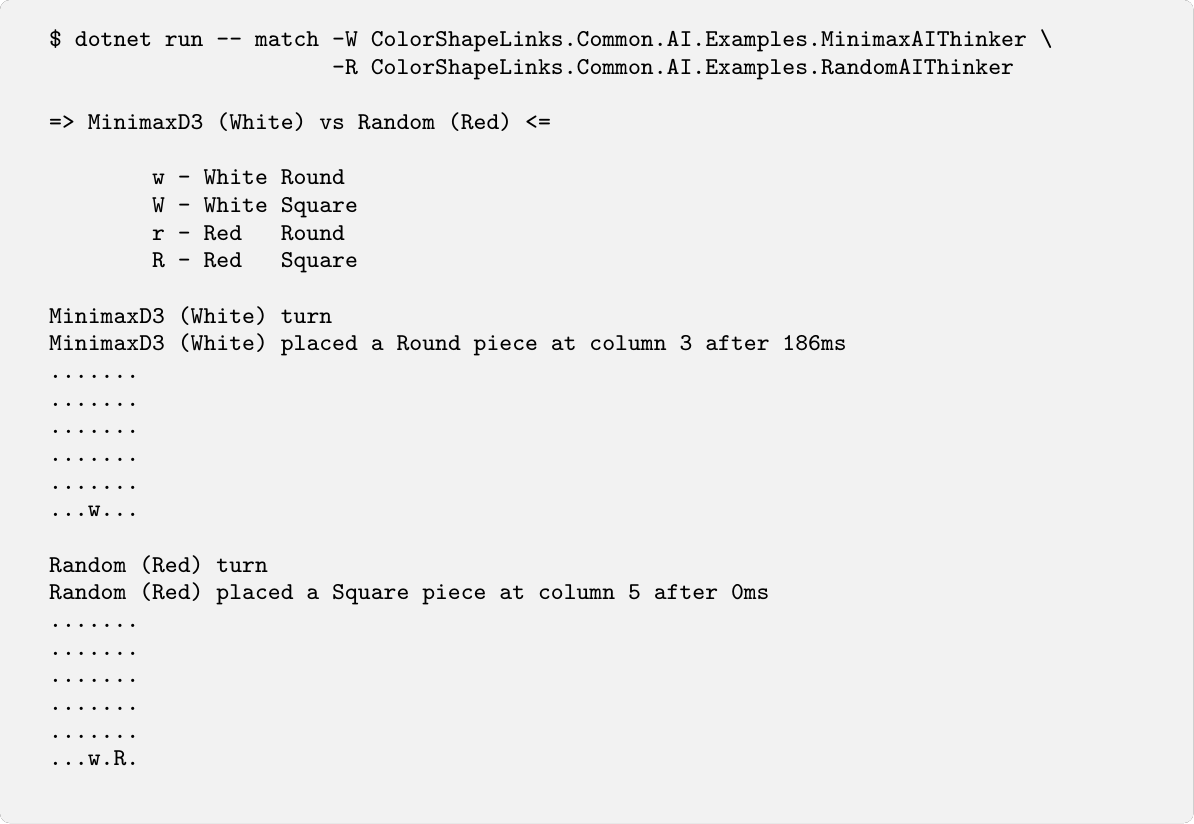}
    \caption{Running ColorShapeLinks using the console UI. Only the start of
        the match is shown.}
    \label{fig:consoleapp}
\end{figure}

The console UI allows for a more refined control of ColorShapeLinks matches
and sessions. Being based on MVC, it allows for easily swapping the UI, as
well as running matches at full speed, contrary to the Unity editor.
\figurename~\ref{fig:consoleapp} shows running a ColorShapeLinks match using
the console UI.

While the console frontend offers many extensibility points, its default
configuration will likely suffice in most situations. For example, users can
run a learning algorithm on top of the console app, making use of its
return values, which indicate the match result or if an error occurred.

\subsection{Implementing an Agent}
\label{sec:devframework:implementing}

The first step to implement an AI agent is to extend the
\texttt{AbstractThinker} base class. This class has three overridable methods,
but it is only mandatory to override one of them, as shown in
Table~\ref{tab:thinkmethods}.

\begin{table}[tb]        \centering    \caption{Overridable Methods in the \texttt{AbstractThinker} class}
    \label{tab:thinkmethods}
    \begin{tabular}{lll}
        \toprule
    Method & Mandatory & Purpose\\
           & override? &        \\
        \midrule
    \texttt{Setup()}    & No  & Setup the AI agent.\\
    \texttt{Think()}    & Yes & Select the next move to perform.\\
    \texttt{ToString()} & No  & Return the AI agent's name.\\
        \bottomrule
    \end{tabular}
\end{table}

There is also the non-overridable \texttt{OnThinkingInfo()} method, which can be
invoked for producing ``thinking'' information, mainly for debugging purposes.
In the Unity frontend this information is printed on Unity's console, while
in the console frontend the information is forwarded to the registered
thinker listeners (or views, from a MVC perspective).

Classes extending \texttt{AbstractThinker} also inherit a number of useful
read-only properties, namely board and match configuration properties (number of
rows, number of columns, number of pieces in sequence to win a game, number
of initial round pieces per player and number of initial square pieces per
player) and the time limit for the AI to play. Concerning the
board/match configuration properties, these are also available in the
board object given as a parameter to the \texttt{Think()} method. However,
the \texttt{Setup()} method can only access them via the inherited properties.

The following subsections address the overriding of each of these three
methods.

\subsubsection{Overriding the \texttt{Setup()} method}
\label{sec:devframework:implementing:setup}

If an AI agent needs to be configured before starting to play, the
\texttt{Setup()} method is the place to do it. This method receives a single
argument, a string, which can contain agent-specific parameters, such
as maximum search depth, heuristic to use, and so on. It is the agent's
responsibility to parse this string. In the Unity frontend, the string is
specified in the ``Thinker params'' field of the \texttt{AIPlayer} component.
When using the console frontend, the string is passed via the
\texttt{-{}-white/red-params} option for simple matches, or after the agent's
fully qualified name in the configuration file of a complete session.
Besides the parameters string, the \texttt{Setup()} method also has access to
board/match properties inherited from the base class.

The same AI agent can represent both players in a single match, as well as more
than one player in sessions/tournaments. Additionally, separate instances of
the same AI agent can be configured with different parameters. In such a
case it might be useful to also override the \texttt{ToString()} method for
discriminating between the instances configured differently. This is an
essential feature if ColorShapeLinks is running under a machine learning and/or
optimization infrastructure.

Note that concrete AI agents require a parameterless constructor in order
to be found by the various frontends. Such constructor exists by default in
C\# classes if no other constructors are defined. However, it is not advisable
to use a parameterless constructor to setup an AI agent, since the various
board/match properties will not be initialized at that time. This is yet
another good reason to perform all agent configuration tasks in the
\texttt{Setup()} method. In any case, concrete AI agents do not need to
provide an implementation of this method if they are not parameterizable or if
they do not require an initial configuration step.

\subsubsection{Overriding the \texttt{Think()} method}
\label{sec:devframework:implementing:think}

The \texttt{Think()} method is where the AI actually does its job and is
the only mandatory override when extending the \texttt{AbstractThinker}
class. This method accepts the game board and a cancellation token,
returning a \texttt{Future\hyp{}Move} object. In other words, the \texttt{Think()}
method accepts the game board, the AI decides the best move to perform, and
returns it. The selected move will eventually be executed by the match
engine.

The \texttt{Think()} method is called in a separate thread. Therefore, it
should only access local instance data. The main thread may ask the AI to
stop thinking, for example if the thinking time limit has expired. Thus,
while thinking, the AI should frequently test if a cancellation request
was made to the cancellation token. If so, it should return immediately
with no move performed.

The game board can be freely modified within the \texttt{Think()} method,
since this is a copy and not the original game board being used in the
main thread. More specifically, the agent can try moves with the
\texttt{DoMove()} method, and cancel them with the \texttt{UndoMove()} method.
The board keeps track of the move history, so the agent can perform any
sequence of moves, and roll them back afterwards. For parallel implementations,
the agent can create additional copies of the board, one per thread, so that
threads can search independently of each other.

The \texttt{CheckWinner()} method of the game board is useful to determine if
there is a winner. If there is one, the solution is placed in the method's
optional parameter.
For building heuristics, the game board's public read-only variable
\texttt{winCorridors} will probably be useful. This variable is a collection
containing all corridors (sequences of positions) where promising or winning
piece sequences may exist.

The AI agent will lose the match in the following situations:

\begin{itemize}
  \item Causes or throws an exception.
  \item Takes too long to play.
  \item Returns an invalid move.
\end{itemize}

\subsubsection{Overriding the \texttt{ToString()} method}
\label{sec:devframework:implementing:tostring}

By default, the \texttt{ToString()} method removes the namespace from the
agent's fully qualified name, as well as the ``thinker'', ``aithinker'' or
``thinkerai'' suffixes. However, this method can be overridden in order to
behave differently. One such case is when agents are parameterizable, and
differentiating between specific parameterizations during matches and
sessions becomes important.

\subsubsection{Summing-up}
\label{sec:devframework:implementing:bottomline}

Building an AI agent for ColorShapeLinks is very simple, asking only the
implementer to extend a class and implement one method. A basic Minimax
algorithm with a simple heuristic can be implemented in less than 30 minutes.
Educators can use it to demonstrate how to create a simple agent from scratch,
during a class for example, and leave up to the students to find better, more
efficient solutions. This can be done as an assignment, a competition, or both.

\subsection{Included Agents}
\label{sec:devframework:agents}

Three test agents are included with the framework, serving both as an example
on how to implement an agent, as well as a baseline reference for testing other
agents.

The \textit{sequential} agent always plays in sequence, from the first to the
last column and going back to the beginning, although skipping full columns. It
will start by using pieces with its winning shape, and when these are over, it
continues by playing pieces with the losing shape. Therefore, it is not a
``real'' AI agent.

The \textit{random} agent plays random valid moves, avoiding full columns and
unavailable shapes. It can be parameterized with a seed to perform
the same sequence of moves in subsequent matches (as long as the same valid
moves are available from match to match).

The \textit{minimax} agent uses a basic, unoptimized Minimax algorithm with a
naive heuristic which privileges center board positions. It can be parameterized
with a search depth, and, although simple, is able to surprise unseasoned
human players---even at low search depths.

\subsection{Availability}
\label{sec:devframework:availability}

The development framework is available at
\url{https://github.com/VideojogosLusofona/color-shape-links-ai-competition}
and is fully open source, licensed under the Mozilla
Public License 2.0\footnote{\url{https://www.mozilla.org/en-US/MPL/2.0/}}
(MPL2), which requires changes to the source code to be shared, although
allowing for integration with proprietary code if the MPL2 licensed code
is kept in separate files. The framework is completely documented and the
documentation is available at
\url{https://videojogoslusofona.github.io/color-shape-links-ai-competition/docs/html/}.

\section{Deployments}
\label{sec:deployments}

The ColorShapeLinks framework has been used to host two internal competitions
(in separate semesters) in an AI for Games course unit at Lusófona University's
Videogames BA, as well as an international AI competition in the IEEE CoG 2020
conference. These deployments are discussed in the next two subsections.

\subsection{Internal Competitions in AI for Games Course Units}
\label{sec:deployments:internal}

ColorShapeLinks was used as an assignment and internal competition in an
AI for Games course unit during the two semesters of the 2019/20 academic year.
In both semesters, the submitted AI agents and associated reports were graded
preliminarily and separately from the final competition. Since students work in
groups of 2 or 3 elements, an individual discussion was also performed. Results
from the internal competition were used to potentially improve the preliminary
grades, not lower them, since the main goal was to motivate students and have an
engaging class where students watched and commented on the performance of each
others' agents in real time. Nonetheless, if an agent was not competent enough
to enter the competition, i.e., it freezed the Unity project or did not respond
to cancellation requests, up to 1 point could be subtracted from the final
grade. No students from the first semester repeated the course during the
second semester.

\begin{figure}[tb]
    \centering
        \includegraphics[width=\linewidth]{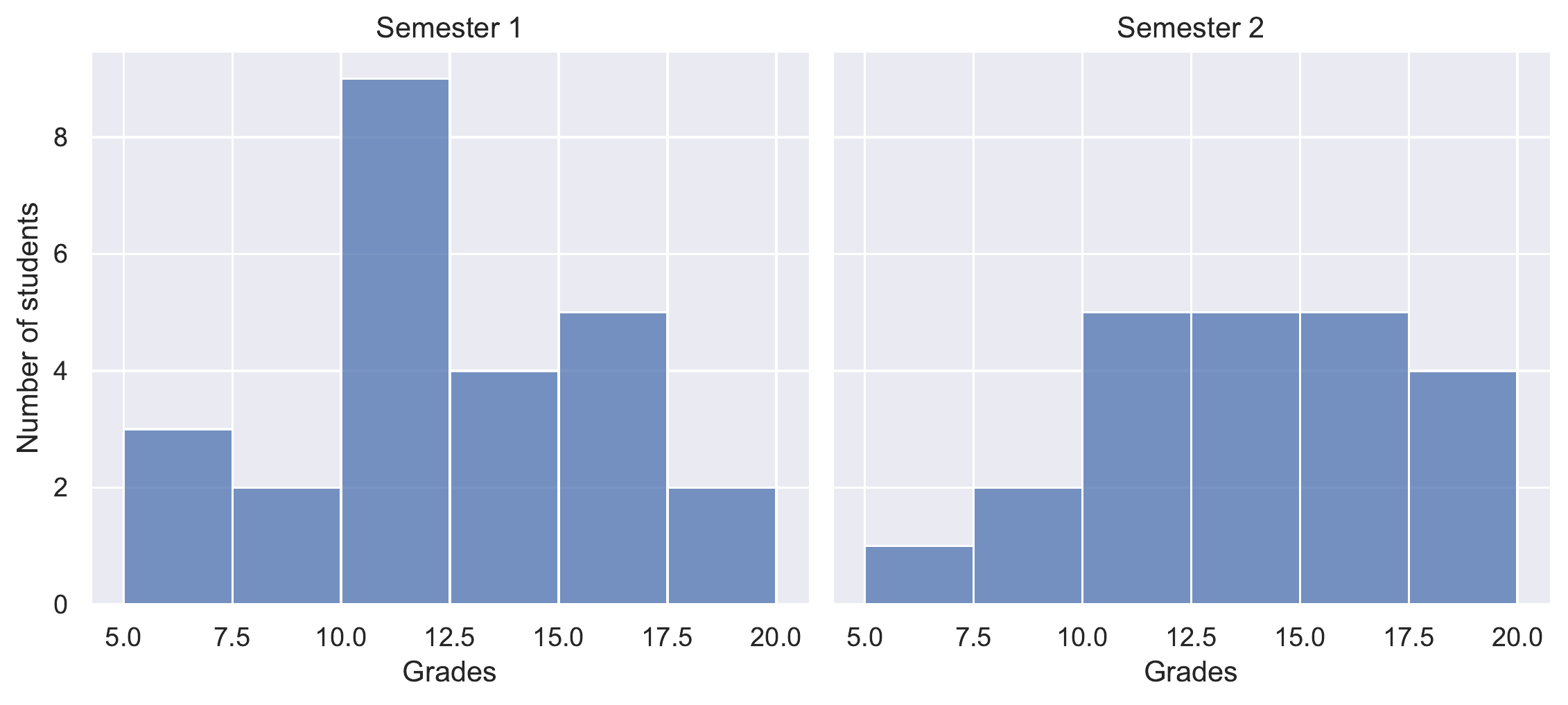}
    \caption{Grade distribution for the ColorShapeLinks AI assignment in the
        two semesters of the 2019/20 academic year.
        Grades are given in a 0--20 scale.}
    \label{fig:grades}
\end{figure}

However, there were several differences between the two semesters. In the first
semester, ColorShapeLinks was a Unity-only assignment, as stated in
Section~\ref{sec:motivation}. The minimum requirement for passing---i.e.,
to have a grade of 10 or higher (grades are given in a 0--20 scale)---was that
students implemented an agent capable of defeating the \textit{sequential} and
\textit{random} agents. The basic \textit{minimax} agent, described in
Subsection~\ref{sec:devframework:agents}, was not included in the framework at
this time. In the second semester, the console frontend and the basic
\textit{minimax} agent were added to the framework, with the assignment
coinciding with the first few weeks of the international competition. The
minimum passing requirement in the second semester was
for the submitted agent to beat the basic \textit{minimax} implementation.
Grade distribution and summary statistics for the ColorShapeLinks assignment in
both semesters are shown in Fig.~\ref{fig:grades} and Table~\ref{tab:grades},
respectively.

\begin{table}
    \caption{\label{tab:grades}Summary statistics for the ColorShapeLinks AI
    assignment grades in the two semesters of the 2019/20 academic year, namely,
    number of students enrolled, $n_\text{all}$, number of students evaluated in
    the assignment, $n_\text{eval}$, mean grade, $\overline{x}$, median grade,
    $\widetilde{x}$, and percentage of students with a passing grade,
    $\%_{\ge 10}$. Grades are given in a 0--20 scale.}
    \centering
    \begin{tabular}{lccccc}
        \toprule
& $n_\text{all}$ & $n_\text{eval}$  & $\overline{x}$ & $\widetilde{x}$ & $\%_{\ge 10}$ \\
        \midrule
     Semester 1 & 27 & 25    & \num{12.02}    & \num{12.00} & \SI{80.00}{\percent} \\
     Semester 2 & 25 & 22    & \num{13.64}    & \num{13.00} & \SI{86.36}{\percent} \\
		\bottomrule
	\end{tabular}
\end{table}

Looking at the overall results in both semesters, at least 80\% of the students
had a passing grade ($\ge 10$) and participated actively in the assignment.
Students less interested in the AI aspect of game design and development mostly
submitted Minimax-based agents, e.g. with alpha-beta pruning, just to pass the
project. They were, nonetheless, generally more committed than in other AI and
programming assignments. On the other end, students more comfortable with
programming and AI generally dedicated more hours to the assignment and came up
with interesting and competitive agents. Most, if not all students were highly
engaged with the final live competition during class (e.g., ``my AI is better
than yours''!).

There were some notable grading variations between the two semesters.
Grades in the second semester were higher in average, and more students met
the passing grade (Table~\ref{tab:grades}), even though the minimum requirement
was higher. Grade distribution (Fig.~\ref{fig:grades}) shows that, in the first
semester, a considerable number of students aimed at the bare minimum just to
get a passing grade, while in the second semester students generally went for
more elaborate solutions. Some of these went beyond the scope of the lectured
materials, making use of transposition tables, search parallelization and
heuristics parameter learning. We argue there were three main reasons,
which separately or in combination, led to this outcome:

\begin{enumerate}
    \item The console app allowed some students to better test and debug
        their agent implementations, without being constrained by Unity.
    \item Some students might have been motivated due to working on an
        assignment which was at the same time an international AI competition.
    \item The \textit{minimax} agent allowed students to study and better
        understand how a ``real'' agent could be implemented, allowing them to
        build upon it.
\end{enumerate}

These results suggest that ColorShapeLinks was successful in allowing students
with various interests and with different capabilities within game making---as
is the case of students in Lusófona's Videogames BA---to
be able to produce palpable, working AI code, and to feel included when
addressing this challenging topic.

\subsection{IEEE CoG 2020 International Competition}

ColorShapeLinks was accepted as an official AI competition at the IEEE CoG 2020
conference, and was funded with a prize money of 500 USD for the winner of each
track. The competition ran on two distinct tracks:

\begin{enumerate}
    \item The \textit{Base Track}, which used standard Simplexity rules with a
        time limit of 0.2 seconds per move. Only one processor core was
        available for the AI agents.
    \item The \textit{Unknown Track}, which was played on a multi-core processor
        under a parameterization known only after the competition deadline,
        since it was dependent on the result of a public lottery draw.
\end{enumerate}

The goal of the \textit{Base Track} was to test agent capabilities in the
standard Simplexity game. The \textit{Unknown Track} evaluated the
generalization power of the submitted solutions when applied to a most likely
untested parameterization.

For each track, the submitted agents played against each other two times, so
each had the opportunity of playing first. Agents were awarded 3 points per
win, 1 point per draw and 0 points per loss, with the final standings for each
track depending on the total number of points obtained per agent.

Classifications for the \textit{Base Track} were updated daily during a five
month period, up until the submission deadline, together with two larger test
parameterizations. This allowed participants to have an idea of how their
submission was faring, and update it accordingly.

The competition had a total of six submissions, four of which were from
undergraduate students, both solo and in teams. Two of the submissions were from
students of the author which fared well in the internal competition discussed
in the previous subsection. Although the number of submissions was low, the fact
that four of them were from undergrads, partially demonstrated that the
competition was accessible.

Eita Aoki, from Japan, won the \textit{Base Track} with his \textit{Thunder}
agent, which used MCTS together with a custom bit board implementation. The
winner and runner-up of the \textit{Unknown Track} were teams from Portugal and
students of the author which attended the AI for Games course in the second
semester. João Moreira won the track with \textit{SureAI}, a highly optimized
Minimax-based agent. The \textit{SimpAI} agent, developed by a
team of three students, was the runner-up, implementing a set of hand-crafted
heuristics balanced with an evolutionary algorithm \citep{fernandes2021simpai}.
While interesting, none of these agents was truly state-of-the-art, leaving the
door open for better agents going forward.

\section{Implications and Limitations}
\label{sec:limitations}

The deployments discussed in the previous sections demonstrate the potential and
usefulness of the ColorShapeLinks framework. The internal competition appeared
to motivate the students, who were highly engaged during the class tournaments
held in both semesters. The introduction of the international competition during
the second semester had a more pronounced effect, though. Overall grades and
dedication improved, with students generally showing more autonomy and going
beyond what was lectured in the classes, effectively validating the works of
\cite{kim2013game} and \cite{yoon2015challenges}. The authors of
\textit{SimpAI}, a student group of the second semester, were able to publish a
paper describing their solution and the second place it obtained in the
\textit{Unknown Track} of the international competition
\citep{fernandes2021simpai}. These results show that AI competitions in general,
and ColorShapeLinks in particular, have clear educational benefits with respect
to student motivation, engagement and autonomy. This is further highlighted by
the fact that Lusófona's Videogames degree is interdisciplinary (non-CS focused)
and industry-driven (not academy-oriented). The fact that students from such a
degree were able to compete in an
international competition with good results indicates that ColorShapeLinks shows
good promise in bridging the gap between game development education and
academic AI.

However, the research presented in this paper has two main limitations which
should be highlighted. First, no survey was done in order to assess students'
reflections on the competitions. The stated student motivation and engagement
during the internal competitions are taken from the author's observations, and
thus, entirely subjective and potentially biased, no matter how clear or obvious
they might have been. Second, the underlying game has a very narrow focus.
Even as an unbounded version of Simplexity, ColorShapeLinks---as a game---is
relatively simple. Thus, many good and excellent solutions are bound to appear
on the short to medium term, possibly even analytical solutions, such as the
case of Connect-4 \citep{allis1988knowledge}. Consequently, ColorShapeLinks---as
a framework---might have a short service life before the problem it addresses
becomes effectively solved. Still, this is not the case at the time of
writing, as ColorShapeLinks will again be one of the competitions being held at
the IEEE CoG 2021 conference.

\section{Conclusions}
\label{sec:conclusions}

In this paper we presented ColorShapeLinks, and AI board game competition
framework specifically designed for game development students and educators,
with openness and accessibility at its core. The arbitrarily-sized Simplexity
board game---implemented by the framework---offers a good balance between
simplicity and complexity, being approachable by undergraduates, while posing a
non-trivial challenge to researchers and more advanced students. The framework
has been successfully used for running internal competitions in an AI for Games
course unit, as well as for hosting an international AI competition, validating
its usefulness. Although the problem addressed by the proposed framework is
bound to be solved in the next few years---thus rendering the framework
obsolete---the general ideas presented in this paper, such as running internal
and international AI competitions using industry standard tools and software
best practices for educational purposes, were confirmed, and remain valid for
similar future endeavors.

\section*{Statements on open data and ethics}

The data reported in this paper is fully anonymized and publicly available at
the Zenodo open-access scientific repository \citep{fachada2021csldata}.

\section*{Declaration of competing interest}

The authors declare that they have no known competing financial interests or
personal relationships that could have appeared to influence the work reported
in this paper.

\section*{Acknowledgments}

This work is supported by Fundação para a Ciência e a Tecnologia under
Grants UIDB/04111/2020 (COPELABS) and UIDB/05380/2020 (HEI-Lab).
The author would like to thank André Fachada for proof-reading the text.
The author would also like to thank the anonymous referees for their valuable
comments and helpful suggestions.

\bibliographystyle{model5-names}
\bibliography{biblio_v2.bib}

\end{document}